\documentclass[a4paper,12pt]{article}
\usepackage[pctex32]{graphics}

\begin{document}
\topmargin 0pt \oddsidemargin 0mm

\renewcommand{\thefootnote}{\fnsymbol{footnote}}
\begin{titlepage}
\begin{flushright}
INJE-TP-07-05\\
\end{flushright}

\vspace{5mm}
\begin{center}
{\Large \bf Instability of  holographic dark energy models}
\vspace{12mm}

{\large   Yun Soo Myung\footnote{e-mail
 address: ysmyung@inje.ac.kr}}
 \\
\vspace{10mm} {\em  Institute of Mathematical Science and School
of Computer Aided Science \\ Inje University, Gimhae 621-749,
Korea }

\end{center}

\vspace{5mm} \centerline{{\bf{Abstract}}}
 \vspace{5mm}
We investigate the difference between holographic dark energy,
Chaplygin gas, and  tachyon model with constant potential. For
this purpose, we examine  their squared speeds of sound  which are
evaluated to zeroth order in perturbation theory and hence depends
only on time. We find that the squared  speed for holographic dark
energy is always negative when choosing  the future event horizon
as the IR cutoff, while those for Chaplygin gas and tachyon are
non-negative. This means that  the perfect fluid for holographic
dark energy is classically unstable. Hence the holographic
interpretation for Chaplygin gas and tachyon is problematic.

\end{titlepage}
\newpage
\renewcommand{\thefootnote}{\arabic{footnote}}
\setcounter{footnote}{0} \setcounter{page}{2}

\section{Introduction}
Observations of supernova type Ia suggest that our universe is
accelerating~\cite{SN}. Considering the ${\rm \Lambda}$CDM
model~\cite{SDSS,Wmap1}, the dark energy and cold dark matter
contribute $\Omega^{\rm ob}_{\rm \Lambda}\simeq 0.74$ and
$\Omega^{\rm ob}_{\rm CDM}\simeq 0.22$ to the critical density of
the present universe. Recently the combination of WMAP3 and
Supernova Legacy Survey data shows a significant constraint on the
equation of state (EOS) for the dark energy, $w_{\rm
ob}=-0.97^{+0.07}_{-0.09}$ in a flat universe~\cite{WMAP3,SSM}.

Although there exist a number of dark energy models~\cite{CST},
the two promising candidates are the cosmological constant  and
the quintessence scenario~\cite{UIS}. The EOS for the latter is
determined dynamically by the scalar or tachyon.

On the other hand, there exists another model of the dark energy
arisen from the holographic principle. The authors in~\cite{CKN}
showed that in quantum field theory, the ultraviolet (UV) cutoff
$\Lambda$ could be related to the infrared (IR) cutoff $L$ due to
the limit set by forming a black hole. If $\rho_{\rm
\Lambda}=\Lambda^4$ is the vacuum energy density caused by the UV
cutoff, the total energy for a system of size $L$ should not
exceed the mass of the system-size black hole:
\begin{equation} \label{1eq1}
E_{\rm \Lambda} \le E_{BH}  \longrightarrow L^3 \rho_{\rm
\Lambda}\le  M_{\rm p}^2L.
\end{equation}
If the largest cutoff $L_{\rm \Lambda}$ is chosen to be the one
saturating this inequality,  the holographic energy density is
given by the energy density of a system-size black hole as
\begin{equation} \label{1eq2}
\rho_{\rm \Lambda}=\frac{3c^2M_{\rm p}^2}{8\pi L_{\rm \Lambda}^2}
\end{equation}
 with a
constant $c$. Here we regard $\rho_{\rm \Lambda}$ as a dynamical
cosmological constant. At the planck scale of $L_{\rm
\Lambda}=M_{\rm p}^{-1}$, it is just the vacuum energy density
$\rho_{\rm V}=M^2_{\rm p}\Lambda_{\rm eff}/8 \pi$ of the universe
at $\Lambda_{\rm eff} \sim M^2_{\rm p}$: $\rho_{\rm \Lambda} \sim
\rho_{\rm p} \sim M^4_{\rm p}$. This implies that a very small
system has an upper limit on the energy density as expected in
quantum field theory. On the other hand, a larger system gets a
smaller energy density. If the IR cutoff is taken as the size of
the current universe ($L_{\rm \Lambda}=H_0^{-1}$), the resulting
energy density is close to the current dark energy: $\rho_{\rm
\Lambda} \sim \rho_{\rm c}\sim 10^{-123}M^4_{\rm p}$~\cite{HMT}.
This results from the holography: the energy increases with the
linear size, so that the energy density decreases with the
inverse-area law. The total energy density dilutes as $L_{\rm
\Lambda}^{-3}$ due to the evolution of the universe, whereas its
upper limit set by gravity (black hole) decreases as $L_{\rm
\Lambda}^{-2}$.

It is not easy to determine the EOS for  a system including
gravity with the UV and IR cutoffs. If one considers $L=H_0^{-1}$
together with the cold dark matter, the EOS may take the form of
$w_{\rm \Lambda}=0$~\cite{HSU}, which is just that of the cold
dark matter. However, introducing an interaction between
holographic dark energy and cold dark matter may lead to an
accelerating universe~\cite{Hor}.  Interestingly, the future event
(particle) horizons\footnote{Here, we introduce the definition of
the future event horizon $R_{\rm
FH}=a(t)\int_{t}^{\infty}\frac{dt'}{a(t')}$ and the particle
horizon $R_{\rm PH}=a(t)\int_{0}^{t}\frac{dt'}{a(t')}$ with the
flat Friedmann-Robertson-Walker metric $ds^2_{\rm
FRW}=-dt^2+a^2(t)d{\bf x}\cdot d{\bf x}$.} were introduced to
obtain the equations of state~\cite{LI,HM,FEH,Myung2,KLM}.

Recently, there was an attempt to make a correspondence between
the holographic dark energy  and Chaplygin gas~\cite{SetaC}. Also
the connection between the holographic dark energy and tachyon
model ~\cite{Sen1,Sen2,Gibb} was introduced to  explain the dark
energy~\cite{SetaT,ZZL}. In the cases of Chaplygin gas with
$p=-A/\rho$~\cite{KMP,FGS} and tachyon model with
$V(T)=\sqrt{A}$~\cite{FKS}, one has the EOS range of $-1\le
\omega_{\rm C,T} \le 0$. Also we have a similar range $-1\le
\omega_{\rm \Lambda} \le -1/3$ for the holographic dark energy
with the future event horizon~\cite{LI}. In spite of the
 similarity between the holographic dark energy
model and Chaplygin gas (tachyon model), there exist differences.
We consider the linear perturbation of holographic dark energy
towards a dark energy-dominated universe. For this purpose, a key
quantity is the squared speed of sound $v^2=d p/d \rho$~\cite{PR}.
The sign of $v^2$ is crucial for determining the stability of a
background evolution. If this is negative, it means a classical
instability of a given perturbation. It is known that the
Chaplygin gas (tachyon) have the positive squared speeds of sound
with $v_{\rm C,T}^2=-\omega_{\rm C,T}$ and thus they are supposed
to be stable against  small perturbations~\cite{GKMPS,STZW}.
Interestingly, the squared speed of sound takes a similar form
like the statefinder parameters $\{r,s\}$~\cite{SSSA}, which can
probe the dynamical evolution of the universe through the higher
derivatives $d^3a/dt^3$ of the scale factor $a$~\cite{Zhang,SZZ}.

In this Letter, we address this issue for the holographic dark
energy model. We compare the holographic dark energy model with
the Chaplygin gas and tachyon model to show its unstable
evolution.

\section{Squared speed for holographic dark energy}
In this section we discuss the flat universe. If the holographic
dark energy density $\rho_{\rm \Lambda}=\frac{3c^2M_{\rm
p}^2}{8\pi L_{\rm \Lambda}^2}$ is known with the IR cutoff $L_{\rm
\Lambda}$, its pressure is determined solely by the conservation
of energy-momentum tensor with $x=\ln a$ ~\cite{HM}
\begin{equation}
p_{\rm \Lambda}=-\frac{1}{3}\frac{d\rho_{\rm \Lambda}}{d
x}-\rho_{\rm \Lambda}\end{equation}
 which  provides the EOS
 \begin{equation}
\omega_{\rm \Lambda}=\frac{p_{\rm \Lambda}}{\rho_{\rm
\Lambda}}=-1+\frac{2}{3H}\frac{\dot{L}_{\rm \Lambda}}{L_{\rm
\Lambda}}.
 \end{equation}
  Hence,
if one does not choose an appropriate form of $L_{\rm \Lambda}$,
one cannot find its EOS. For example, if one chooses the Hubble
horizon $L_{\rm \Lambda}=1/H_0$, it does not provide the correct
EOS~\cite{HSU}. Here we have
$\dot{H}=-\frac{3}{2}H^2(1+\omega_{\rm \Lambda})$, which is
nothing but the second Friedmann equation. On the other hand,
choosing $L_{\rm \Lambda}=R_{\rm
PH/FH}=c/H\sqrt{\Omega_{\Lambda}}$ with $\Omega_{\rm \Lambda}=8
\pi \rho_{\rm \Lambda}/3M^2_pH^2$ leads to
\begin{equation}
\omega^{\rm PH/FH}_{\rm \Lambda}=-\frac{1}{3}\Big(1\mp
\frac{2}{c}\sqrt{\Omega_{\rm \Lambda}}\Big)
 \end{equation}
 because the definition of the holographic dark energy density
 implies
\begin{equation}
\dot{\rho}_{\rm \Lambda}=2H \rho_{\rm \Lambda}\Big[-1 \mp
\frac{1}{HR_{\rm PH/FH}}\Big]=-3H\rho_{\rm
\Lambda}\Big[1-\frac{1}{3}\pm\frac{2\sqrt{\Omega_{\rm
\Lambda}}}{3c}\Big].\end{equation} $\omega_{\rm \Lambda}$ is
determined by the  evolution equation
\begin{equation}
\frac{d\Omega_{\rm \Lambda}}{dx}=\frac{\dot{\Omega}_{\rm
\Lambda}}{H}=-3\omega_{\rm \Lambda}\Omega_{\rm
\Lambda}(1-\Omega_{\rm \Lambda}).
\end{equation}
For our purpose, we introduce the squared speed of holographic
dark energy fluid as
\begin{equation}
v^2_{\rm \Lambda}=\frac{dp_{\rm \Lambda}}{d\rho_{\rm
\Lambda}}=\frac{\dot{p}_{\rm \Lambda}}{\dot{\rho}_{\rm \Lambda}},
\end{equation}
where
\begin{equation}
\dot{p}_{\rm \Lambda}={\dot{\omega}_{\rm \Lambda}}\rho_{\rm
\Lambda}+\omega_{\rm \Lambda}\dot{\rho}_{\rm \Lambda}
\end{equation}
with~\cite{Zhang}
\begin{equation}
\dot{\omega}_{\rm \Lambda}=H \frac{d \omega_{\rm \Lambda}}{dx}=
-\frac{H}{3c}\sqrt{\Omega_{\rm \Lambda}}(1-\Omega_{\rm
\Lambda})\Big(1\mp\frac{2}{c}\sqrt{\Omega_{\rm \Lambda}}\Big).
\end{equation}
 It leads to
\begin{equation}
v^2_{\rm \Lambda}=\omega_{\rm
\Lambda}\Big[1-\frac{\sqrt{\Omega_{\rm \Lambda}}(1-\Omega_{\rm
\Lambda})}{3c(1+\omega_{\rm \Lambda})}\Big]=\omega_{\rm
\Lambda}\Big[1-\frac{\sqrt{\Omega_{\rm \Lambda}}(1-\Omega_{\rm
\Lambda})}{2(c\pm\sqrt{\Omega_{\rm \Lambda}})}\Big],
\end{equation}
which contrasts to those for the Chaplygin gas and tachyon model
\begin{equation}
v^2_{\rm C,T}=-\omega_{\rm C,T} \ge 0.
\end{equation}
In the linear perturbation theory, the density perturbation is
described by
\begin{equation}
\rho(t,{\bf x})=\rho(t)+\delta\rho(t,{\bf x})
\end{equation}
with $ \rho(t)$ the background value. Then the conservation law
for the energy-momentum tensor of $\nabla_{\nu}T^{\mu\nu}=0$
yields~\cite{KimHS}
\begin{equation} \label{pcl}
\delta \ddot{\rho}=v^2 \nabla^2 \delta \rho(t,{\bf x}),
\end{equation}
where $T^0~_0=-(\rho(t)+\delta\rho(t,{\bf x}))$ and
$v^2=dp/d\rho$. For $v^2_{\rm C,T}>0$, Eq. (\ref{pcl}) becomes a
regular wave equation whose solution is given by $\delta \rho_{\rm
C,T}=\delta\rho_{\rm 0 C,T} e^{-i\omega t+i {\bf k}\cdot {\bf
x}}$. Hence the positive squared speed (real value of speed) shows
a regular propagating mode for a density perturbation. For
$v^2_{\rm \Lambda}<0$, the perturbation becomes an irregular wave
equation whose solution is given by $\delta \rho_{\rm
\Lambda}=\delta\rho_{\rm 0 \Lambda} e^{\omega t+i {\bf k}\cdot
{\bf x}}$. Hence the negative squared speed (imaginary value of
speed) shows an exponentially growing mode for a density
perturbation. That is, an increasing density perturbation induces
a lowering pressure, supporting the emergence of instability. In
Table 1, we summarize the relevant quantities for holographic dark
energy, Chaplygin gas, and tachyon model for comparison.

\begin{table}
\caption{Summary for holographic dark energy (HDE), Chaplygin gas
 (CG), tachyon model (TM). For HDE, the conservation law determines its pressure because the energy density
 is known, while for CG, the conservation law determines the energy density because the pressure is known.
 Range of EOS for HDE is for the future event horizon. }
\begin{tabular}{|c|c|c|c|}
  \hline
  % after \\: \hline or \cline{col1-col2} \cline{col3-col4} ...
   & HDE & CG & TM \\
  \hline
  energy density & $\rho_{\rm \Lambda}=3c^2M_{\rm p}^2/8 \pi L^2_{\rm \Lambda}$ & $\rho_{\rm C}=\sqrt{A+B/a^6}$
  & $\rho_{\rm T}=V/\sqrt{1-\dot{T}^2}$ \\
  pressure &$p_{\rm \Lambda}=\omega_{\rm \Lambda} \rho_{\rm \Lambda}$ & $p_{\rm C}=-A/\rho_{\rm C}$
  & $p_{\rm T}=-V\sqrt{1-\dot{T}^2}$ \\
  EOS & $\omega_{\rm \Lambda}=-1/3 \pm 2 \sqrt{\Omega_{\rm \Lambda}}/3c$ & $\omega_{\rm C}=-A/\rho_{\rm C}^2$& $\omega_{\rm T}=-1+\dot{T}^2$ \\
  range of EOS& $-1 \le \omega_{\rm \Lambda} \le -1/3$ & $-1 \le \omega_{\rm C} \le 0 $ & $-1 \le \omega_{\rm T} \le 0$  \\
  squared speed& $v^2_{\rm \Lambda}=\omega_{\rm
\Lambda}\Big[1-\frac{\sqrt{\Omega_{\rm \Lambda}}(1-\Omega_{\rm
\Lambda})}{3c(1+\omega_{\rm \Lambda})}\Big]$ & $v^2_{\rm C}=A/\rho^2=-\omega_{\rm C}$ &  $v^2_{\rm T}=1-\dot{T}^2=-\omega_{\rm T}$\\
  \hline
\end{tabular}
\end{table}

 In the case of holographic dark energy with the future
event horizon, one finds from Fig. 1 that the squared speed is
always negative for the whole evolution $0 \le \Omega_{\rm
\Lambda} \le 1$. Especially, for $c=0.8(<1)$, we have a
discontinuity from $v^2_{\rm \Lambda}=-\infty$ to $\infty$ around
$\Omega_{\rm \Lambda}=0.64$ whose equation of state crosses
$\omega_{\rm \Lambda}=-1$. For example, we have $v^2_{\rm
\Lambda}=229$ at $\Omega_{\rm \Lambda}=0.639$ while we have
$v^2_{\rm \Lambda}=-231$ at $\Omega_{\rm \Lambda}=0.641$. This
means that the phantom phase occurs when the squared speed of
holographic dark energy blows up.
\begin{figure}[t!]
\centering
   {\includegraphics{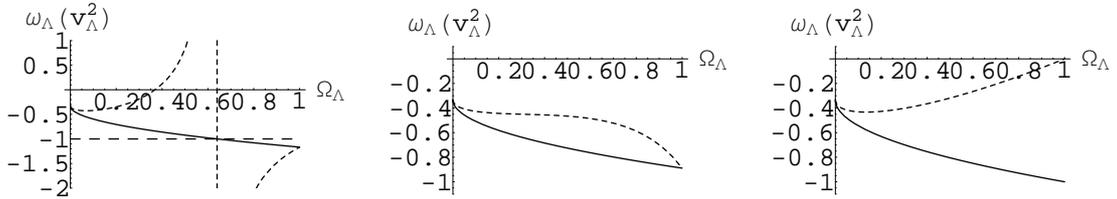}}
\caption{ Three graphs for the holographic dark energy with the
future event horizon. The solid (dashed) lines denote the equation
of state $\omega_{\rm \Lambda}$ (squared speed $v^2_{\rm
\Lambda}$). One has the graphs for $c=0.8$, $c=1$, and $c=1.2$
from the left to the right. } \label{fig1}
\end{figure}

In the case of holographic dark energy with the particle horizon,
one finds from Fig. 2 that the $c=1$ squared speed changes from
$-1/3$ to 1/3 as the universe evolves,  which is nearly coherent
with the equation of state. Also there is no sizable difference
between $c=0.8,1.0$ and 1.2 except  slightly different loci for
$v^2_{\rm \Lambda}=0$. In this case,  we read off the classical
instability of $v^2_{\rm \Lambda}<0$ for $-1/3 \le \omega_{\rm
\Lambda}<0$.
\begin{figure}[t!]
\centering
   {\includegraphics{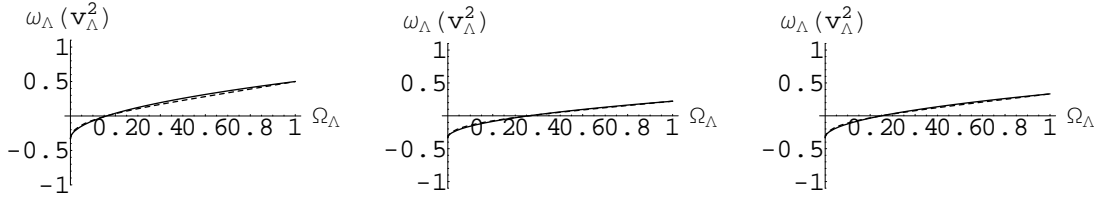}}
\caption{Three graphs for the holographic dark energy with the
particle horizon. The solid (dashed) lines denote the equation of
state $\omega_{\rm \Lambda}$ (squared speed $v^2_{\rm \Lambda}$).
From the left to the right, one has the graphs for $c=0.8$, $c=1$,
and $c=1.2$. There is no significant change between them. }
\label{fig2}
\end{figure}
However, from Fig. 3 we have  the non-negative squared speed in
the Chaplygin gas model. This means that the Chaplygin gas is
stable against the linear perturbation even though it could
describe both the cold dark matter at the early universe and dark
energy at the present and future universe. Also for the tachyon
model with constant potential which is essentially the Chaplygin
gas, we have the cold dark matter at $\dot{T}=1~(T\to \infty,
\ddot{T}=0)$ and dark energy at $\dot{T}=0~(T \simeq {\rm const},
\ddot{T}=0)$. On the other hand, for $V(T) \simeq m^2(T-T_0)^2/2$
, fluctuations coupled to the oscillating background condensate
were exponentially unstable and for pressureless tachyon with
$V(T)=V_0e^{-T/T_0}$, fluctuations coupled with metric
perturbations   also showed gravitational instability~\cite{FKS}.

\begin{figure}[t!]
\centering
   {\includegraphics{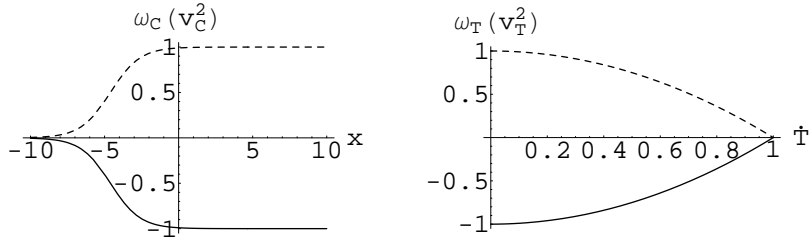}}
\caption{Two graphs for the Chaplygin gas and tachyon model. The
left panel is for $\omega_{\rm C}(v^2_{\rm C})$ vs $x=\ln a$,
while the right one is for  $\omega_{\rm T}~(v^2_{\rm T})$ vs
$\dot{T}$. The solid (dashed) lines denote the equation of state
$\omega_{\rm C,T}$(squared speed $v^2_{\rm C,T}$). Here we find
the positive squared speeds. } \label{fig3}
\end{figure}

\section{Squared speed for the nonflat universe}
In this section, we attempt to find the squared speed for the
nonflat universe. For this purpose, we introduce the density
parameter for curvature defined by $\Omega_{\rm
k}=\frac{k}{a^2H^2}$. Then we can  rewrite the Friedmann equation
as a simplified form

\begin{equation} \label{3eq2} \Omega_{\rm
\Lambda}=1+\Omega_{\rm k}.\end{equation} For the non-flat universe
of $k \not=0$, we  consider the future event horizon $L_{\rm
\Lambda}=R_{\rm FH}=a\xi_{\rm FH}(t)=a \xi^{k}_{\rm FH}(t)$ with
\begin{equation}
\label{3eq3} \xi_{\rm FH}(t)=\int_t^{\infty} \frac{dt}{a}.
\end{equation}
Here the comoving horizon size is given by
\begin{equation} \label{3eq4}
\xi^k_{\rm
FH}(t)=\int_{0}^{r(t)}\frac{dr}{\sqrt{1-kr^2}}=\frac{1}{\sqrt{|k|}}{\rm
sinn}^{-1}\Bigg[\sqrt{|k|}r(t)\Bigg],
\end{equation}
which  leads to $\xi^{k=1}_{\rm FH}(t)={\rm sin}^{-1}r(t)$,
$\xi^{k=0}_{\rm FH}(t)=r(t)$, and  $\xi^{k=-1}_{\rm FH}(t)={\rm
sinh}^{-1}r(t)$. Here we  introduce  a comoving radial coordinate
$r(t)$,
\begin{equation} \label{3eq5}
r(t)=\frac{1}{\sqrt{|k|}} {\rm sinn}\Bigg[\sqrt{|k|}\xi^{k}_{\rm
FH}(t)\Bigg].
\end{equation}
Then $L_{\rm \Lambda}=ar(t)$ is a useful length scale  for the
non-flat universe~\cite{HM}. Its derivative with respect to time
$t$ leads to
\begin{equation} \label{3eq6}
\dot{L}_{\rm \Lambda}=H L_{\rm \Lambda}+a
\dot{r}=\frac{c}{\sqrt{\Omega_{\rm \Lambda}}}-{\rm cosn}y,
\end{equation}
where ${\rm cosn}y={\rm cos}y,~y,~{\rm cosh}y$ for $k=1,0,-1$ with
$y=\sqrt{k} R_{\rm FH} /a$.  One finds the  equation of state for
the holographic dark energy
\begin{equation} \label{3eq7}
\dot{\rho}_{\rm \Lambda} +3H \Big[1-
\frac{1}{3}-\frac{2\sqrt{\Omega_{\rm \Lambda}}}{3c}{\rm
cosn}y\Big]\rho_{\rm \Lambda}=0.
\end{equation}
Here we can read off the  EOS
\begin{equation}
\label{3eq8} \omega_{\rm \Lambda}=
-\frac{1}{3}-\frac{2\sqrt{\Omega_{\rm \Lambda}-c^2\Omega_{\rm
k}}}{3c}=-\frac{1}{3}-\frac{2\sqrt{\Omega_{\rm
\Lambda}(1-c^2)+c^2}}{3c}.
\end{equation}
In this case, considering  the evolution equation together with
Eq.(\ref{3eq2}) leads to
\begin{equation}
\frac{d\Omega_{\rm \Lambda}}{dx}=-3\omega_{\rm \Lambda}\Omega_{\rm
\Lambda}(1-\Omega_{\rm \Lambda})+\Omega_{\rm \Lambda}\Omega_{\rm
k}=-(3\omega_{\rm \Lambda}-1)\Omega_{\rm \Lambda}(1-\Omega_{\rm
\Lambda}).
\end{equation}
Finally, the squared speed takes the form

\begin{equation} v^2_{\rm \Lambda}=\frac{\dot{p}_{\rm \Lambda}}{\dot{\rho}_{\rm \Lambda}}=\omega_{\rm
\Lambda}-\frac{(1-c^2)\Omega_{\rm \Lambda}(1-\Omega_{\rm
\Lambda})(c+\sqrt{\Omega_{\rm \Lambda}(1-c^2)+c^2}
)}{3c(c-\sqrt{\Omega_{\rm \Lambda}(1-c^2)+c^2})\sqrt{\Omega_{\rm
\Lambda}(1-c^2)+c^2}}.
\end{equation}
 In the case of the nonflat universe  with the future
event horizon, one finds from Fig. 4 that the squared speed
changes from positive value to negative one, as the universe
evolves. Here one has $v^2_{\rm \Lambda}=0$ at $\Omega_{\rm
\Lambda}=0.22(0.27)$ for $c=0.8(1.2)$. For $c=0.8$, there is no
discontinuity. In this sense,  we insist that the nonflat effect
improves the instability of the flat universe when comparing Fig.4
with Fig.1. However, the perfect fluid of holographic dark energy
is still unstable for the nonflat universe.  Furthermore, for
$c=1$, we have de Sitter spacetime of $\omega_{\rm \Lambda}=-1$.
It implies that $\dot{\rho}_{\rm \Lambda}=0$. Hence it is
difficult to define its squared speed for de Sitter spacetime.
\begin{figure}[t!]
\centering
   {\includegraphics{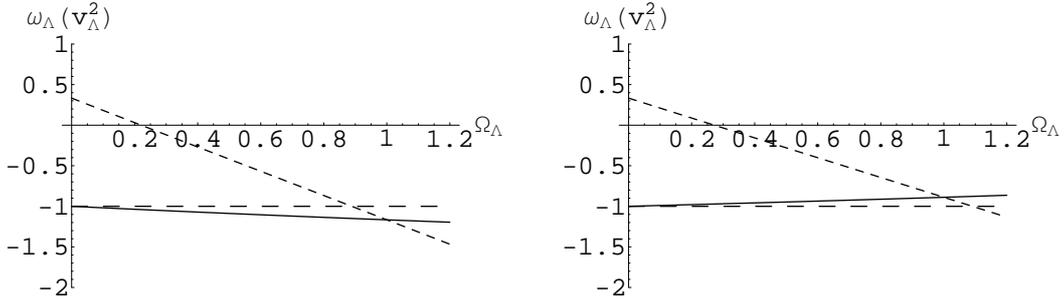}}
\caption{Two graphs for the holographic dark energy with the
future event horizon for $k\not=0$. The solid (dashed) lines
denote the equation of state $\omega_{\rm \Lambda}$ (squared speed
$v^2_{\rm \Lambda}$). From the left to the right, one has the
graphs for $c=0.8$ and $c=1.2$. The horizontal line denotes $-1$.
For $c=1$ case, we cannot define its squared speed because of
$\omega_{\rm \Lambda}=-1$.  } \label{fig4}
\end{figure}

\section{Discussions}
We study the difference between holographic dark energy, Chaplygin
gas, and  tachyon model with constant potential. Especially, we
calculate their squared  speeds which are crucially important to
determine the stability of perturbations. We find that the squared
speed for holographic dark energy is always negative when imposing
the future event horizon as the IR cutoff, while those for
Chaplygin gas and tachyon are always non-negative. This means that
the perfect fluid model  for holographic dark energy is
classically unstable. Hence the holographic interpretation for
Chaplygin gas and tachyon is problematic. Particularly, the
holographic embeddings for Chaplygin gas and
tachyon~\cite{SetaC,SetaT,ZZL} are not guaranteed even though they
have similar equations of state like  the holographic dark energy.

Despite the success of holographic dark energy in obtaining an
accelerating universe for $L_{\rm \Lambda}=R_{\rm FH}$, it may not
give us a promising solution to the dark energy-dominated universe
because choosing the future event horizon just means an unstable
evolution. This contrasts to those for the Chaplygin gas and
tachyon with constant potential.

\section*{Acknowledgment}
 This work
was in part supported by the Korea Research Foundation
(KRF-2006-311-C00249) funded by the Korea Government (MOEHRD) and
the SRC Program of the KOSEF through the Center for Quantum
Spacetime (CQUeST) of Sogang University with grant number
R11-2005-021.

        \end{document}